%% file: main.tex
\documentclass[12pt]{article}

\usepackage[margin=1in]{geometry}
\usepackage{setspace}
\usepackage{mathptmx}          
\usepackage{amsmath}
\usepackage{graphicx}
\usepackage{booktabs}
\usepackage{array}
\usepackage{longtable}
\usepackage{makecell}
\usepackage[round,authoryear]{natbib}
\usepackage[
  colorlinks=true,
  linkcolor=black,
  citecolor=black,
  urlcolor=black
]{hyperref}
\usepackage{microtype}         

\title{\textbf{Private Credit Markets Theory, Evidence, and Emerging Frontiers}}

\author{Jiacheng Zou\thanks{ jiachengzou@alumni.stanford.edu }}

\date{\today}

\begin{document}

\maketitle
\thispagestyle{empty}

\begin{abstract}
\noindent
Private credit assets under management grew from \$158 billion in 2010 to nearly \$2 trillion
globally by mid-2024, fundamentally reshaping corporate credit markets. This paper provides a
systematic survey of the academic literature on private credit, organizing theory and evidence
around four questions: why the market has grown so rapidly, how direct lender technology differs
from bank lending, what risk-adjusted returns investors earn, and whether the sector poses
systemic risks. We develop an integrated theoretical framework linking delegated monitoring,
soft-information processing, and incomplete contracting to the institutional specifics of modern
direct lending. The empirical evidence documents a distinctive lending technology serving opaque,
private-equity-sponsored borrowers at a meaningful and persistent spread premium over the broadly
syndicated loan market, while performance evidence suggests that risk-adjusted returns for the
average fund are largely consumed by fees.
\end{abstract}



\clearpage
\setcounter{page}{1}

\input{sections/01_introduction}

\input{sections/02_market_overview}

\input{sections/03_theory}

\input{sections/04_empirical_evidence}

\input{sections/05_contract_design}

\input{sections/06_fund_structures}

\input{sections/07_systemic_risk}

\input{sections/08_conclusion}

\clearpage
\singlespacing
\bibliographystyle{chicago}
\bibliography{bibliography}

\clearpage
\doublespacing
\input{tables/table1_market_growth}

\clearpage
\input{tables/table2_spread_comparison}

\clearpage
\input{tables/table3_anchor_papers}

\end{document}

%% file: sections/01_introduction.tex

\section{Introduction}
\label{sec:introduction}

In 2010, global private credit assets under management stood at approximately \$158 billion, a
modest corner of the fixed-income landscape dwarfed by syndicated leveraged loan markets and
public high-yield bond issuance. By mid-2024, U.S.\ private credit assets reached approximately
\$1.34 trillion, with global assets approaching \$2 trillion under the Federal Reserve's
measurement framework \citep{BerrospideEtal2025}---a more than eight-fold increase in U.S.\
assets over fourteen years. Using a broader definitional perimeter that includes infrastructure
debt and broadly syndicated loan participations, industry estimates place the global figure at
\$2.1 trillion by end-2023, a 13.3-fold expansion from the 2010 base representing a compound
annual growth rate of 22 percent \citep{Preqin2024,IMF_GFSR2024}. No other major asset class in the post-Global Financial Crisis era has sustained growth of this
magnitude over a comparable horizon. The expansion has transformed the architecture of corporate
credit markets, repositioned the boundary between bank and non-bank financial intermediation, and
raised fundamental questions about the efficiency, stability, and regulatory design of modern
capital markets.

This paper provides the first comprehensive survey of the academic literature on private
credit---defined as non-bank, non-publicly-traded debt financing extended to corporations by
specialized fund vehicles such as business development companies, closed-end funds, and separately
managed accounts. We organize the rapidly growing body of theoretical and empirical work around
four interconnected research questions.

Why has private credit grown so rapidly, and what economic forces---regulatory,
technological, and institutional---have driven the displacement of bank balance-sheet lending by
direct lenders? The answer to this question bears directly on whether the growth of private credit
reflects efficient reallocation of intermediation functions or, instead, regulatory arbitrage that
merely shifts risk from regulated to unregulated balance sheets.
\citet{ChernenkoErelPrilmeier2022} attribute two-thirds of nonbank lending to bank regulatory
constraints, while \citet{ChernenkoIalentiScharfstein2024} find that BDCs hold substantial excess
capital even under severely adverse stress scenarios, complicating a pure regulatory-arbitrage
interpretation. The tension between these findings remains unresolved.

How does the lending technology of direct lenders differ from that of banks, and what are
the implications for loan pricing, contract design, and borrower outcomes?
\citet{JangKimSufi2025} document that direct lenders serve younger, intangible-capital-intensive,
private-equity-sponsored firms, employ blanket liens as collateral in 79 percent of loans
(compared with fewer than 15 percent for banks), and specialize by industry rather than
geography. These patterns suggest a distinctive model of credit provision grounded in
soft-information processing and enterprise-level risk assessment, consistent with the theoretical
predictions of \citet{Stein2002} and \citet{Diamond1984}.

What are the risk-adjusted returns to private credit investors, and do fund structures
create or destroy value? Headline returns appear attractive relative to public credit
alternatives, but \citet{ErelFlanaganWeisbach2024} show that rigorous risk adjustment
substantially reduces apparent outperformance, and that the typical fee structure absorbs much of
the gross alpha that information production and monitoring generate. Whether investors are
adequately compensated for the illiquidity and complexity they bear is among the most
consequential open questions in the field.

Finally, does the growth of private credit pose systemic risks, and how should prudential
regulation adapt? The \citet{IMF_GFSR2024} identifies five key vulnerabilities: fragile
borrowers, semi-liquid investment vehicles, multiple layers of leverage, stale and subjective
valuations, and unclear interconnections among market participants.
\citet{AramonteShrimpfShin2022} develop a theoretical framework in which non-bank leverage
amplifies stress through margin spirals, while the \citet{FederalReserve_PrivateCredit2024}
documents that the floating-rate character of direct lending loans creates a powerful channel for
monetary policy transmission that operates outside the traditional banking system.

The review makes four contributions. We construct an integrated theoretical framework that maps
the classical literatures on delegated monitoring \citep{Diamond1984}, soft-information processing
\citep{Stein2002}, covenant design \citep{RajanWinton1995}, and debt-source sorting
\citep{DenisMihov2003} to the institutional specifics of modern direct lending; provide a
comprehensive synthesis of the post-2020 wave of empirical papers on private credit that have not
yet been organized into a coherent narrative; assess the methodological frontier for measuring
private debt fund performance, identifying the challenges that distinguish private credit from the
better-studied private equity context; and map the systemic-risk channels that connect private
credit to the broader financial system, integrating academic and policy literatures into a single
analytical framework.

The remainder of this paper proceeds as follows. Section~\ref{sec:market_overview} defines the
private credit market, documents its growth trajectory, and describes the structural shifts that
have shaped its evolution. Section~\ref{sec:theory} develops the theoretical foundations, drawing
on information economics, delegated monitoring, and incomplete-contracts theory.
Section~\ref{sec:empirical} reviews the empirical evidence on loan pricing, default, recovery,
and credit quality. Section~\ref{sec:covenants} examines covenant design and lender monitoring in
private credit transactions. Section~\ref{sec:fund_structures} analyzes fund structures,
incentive alignment, and performance measurement. Section~\ref{sec:systemic_risk} addresses
systemic risk and the regulatory implications of non-bank credit expansion.
Section~\ref{sec:conclusion} concludes with a synthesis of the review's findings and identifies
the most promising directions for future research.

%% file: sections/02_market_overview.tex

\section{The Rise of Private Credit Markets}
\label{sec:market_overview}

The institutional landscape of corporate credit provision has undergone a structural
transformation over the past fifteen years. What was once a modest niche occupied by mezzanine
funds and specialty finance companies has grown into a market with approximately \$1.34 trillion
in U.S.\ assets and nearly \$2 trillion globally as of mid-2024 \citep{BerrospideEtal2025},
rivaling the broadly syndicated leveraged loan market in scale and increasingly competing with
it for the same borrowers. Understanding the contours of this transformation---the instruments, the
participants, the growth dynamics, and the regulatory catalysts---is a prerequisite for the
theoretical and empirical analysis that follows. This section defines the scope of private credit,
traces its expansion, and identifies the supply-side and demand-side forces that have driven it.

\subsection{Defining Private Credit and Its Principal Forms}
\label{sec:market_overview:forms}

The term ``private credit'' encompasses a heterogeneous set of non-bank, non-publicly-traded debt
instruments extended to corporate borrowers, typically through specialized fund vehicles. We adopt
the definition used by \citet{IMF_GFSR2024}, which includes direct lending, mezzanine and
subordinated debt, distressed and special-situations debt, and venture lending, while excluding
consumer credit, trade credit, and real-assets debt. This definition aligns closely with the scope
used by \citet{ChernenkoErelPrilmeier2022} and by \citet{BlockJangKaplanSchulze2024} in their
survey of private debt fund managers.

\textbf{Direct lending} constitutes the largest and fastest-growing segment. Direct lending funds
originate term loans---typically unitranche, first-lien, or second-lien facilities---to
middle-market firms, defined loosely as companies with EBITDA between \$10 million and \$100
million. As Table~\ref{tab:market_growth} shows, US middle-market direct lending volume grew from
\$12.8 billion in 2010 to \$177.6 billion in 2023, a compound annual growth rate of 22.4 percent
(calculated as $(177.6/12.8)^{1/13}-1$).
The unitranche structure, which combines senior and subordinated debt into a single facility with
a blended interest rate, has become the dominant form: \citet{BlockJangKaplanSchulze2024} report
that the majority of surveyed GPs originate primarily cash-flow-based unitranche loans. The
borrowers are predominantly private-equity-sponsored, and the loans are secured by blanket liens
on all assets of the borrower and its subsidiaries \citep{JangKimSufi2025}.

\textbf{Mezzanine and subordinated debt} occupies a junior position in the capital structure,
typically subordinated to senior secured term loans and carrying higher coupons with equity-linked
features such as warrants or payment-in-kind toggles. This segment was historically the core of
the private credit market before direct lending displaced it in volume. Mezzanine lending serves
an important gap-filling function in leveraged buyout financing, providing the incremental
leverage that permits private equity sponsors to achieve target equity returns without exceeding
first-lien lenders' risk tolerances.

\textbf{Distressed and special-situations debt} involves the acquisition of discounted claims on
financially impaired borrowers, often with the intent of influencing restructuring outcomes. While
conceptually distinct from origination-focused direct lending, distressed strategies share the
information-intensive, relationship-driven characteristics of private credit more broadly.

\textbf{Venture lending} provides debt financing to venture-capital-backed companies, typically
structured as term loans with warrant coverage. This segment remains small relative to direct
lending but has grown in tandem with the venture capital market.

The boundary between private credit and adjacent asset classes is porous and requires explicit
delineation. Syndicated leveraged loans, originated by banks and distributed to institutional
investors through a syndication process, share many economic features with direct lending but
differ in their origination process, secondary liquidity, and covenant structures.
\citet{BuchakMatvosPiskorskiSeru2024} develop a framework for analyzing the substitution margin
between bank-originated and shadow-bank-originated credit, demonstrating that shadow banks
substitute for traditional banks primarily among loans that are easily sold---a pattern that
highlights the importance of secondary market liquidity in distinguishing these segments. Public
high-yield bonds, while also serving non-investment-grade borrowers, differ from private credit in
their standardized indenture provisions, public registration requirements, and exchange-based
secondary trading. Bank bilateral loans held on balance sheet represent the historical antecedent
of private credit, and the migration of lending activity from bank balance sheets to non-bank fund
vehicles is the central structural shift that this review documents.

\subsection{Market Size, Growth, and Structural Shifts}
\label{sec:market_overview:growth}

The aggregate growth of private credit is the headline empirical fact of this survey. As
Table~\ref{tab:market_growth} documents, global private credit AUM rose from \$158 billion in
2010 to approximately \$1.34 trillion in U.S.\ assets and nearly \$2 trillion globally by
mid-2024---representing roughly five-fold growth since 2009 \citep{BerrospideEtal2025}.
Industry estimates using broader definitional boundaries reach \$2.1 trillion globally for
2023 \citep{Preqin2024}, with the difference reflecting whether infrastructure debt, real
asset lending, and broadly syndicated participations are included. This sustained growth rate
far exceeded the approximately 10 percent CAGR of global private equity over the same period.
Growth accelerated markedly after 2013, coinciding with the implementation of post-Dodd-Frank
leveraged lending guidance by the OCC, FDIC, and Federal Reserve that constrained banks from
originating loans above six times EBITDA leverage.

\textbf{Bank credit to private credit vehicles.} A previously underexamined dimension of
private credit's growth is its reliance on bank funding. Using Federal Reserve supervisory data
(FR~Y-14Q) from the largest U.S.\ bank holding companies, \citet{BerrospideEtal2025} document
that bank committed credit lines to private credit vehicles---BDCs and private debt funds---grew
from approximately \$8 billion in 2013 to \$95 billion by late 2024, with utilized amounts
reaching \$56 billion as of December 2024. This 145 percent growth over five years outpaced both
general nonbank financial institution commitments (53 percent) and nonfinancial corporate lending
(14 percent) over the same period. The major U.S.\ bank lenders are JP~Morgan Chase, Citigroup,
Wells Fargo, and Bank of America; U.S.\ banks serve as lead arrangers on approximately 50 percent
of identified BDC credit facilities, with foreign banks providing another 30 percent. These
linkages imply that bank balance sheets remain materially exposed to private credit performance
through indirect channels even as direct lending has migrated off bank balance sheets.

\textbf{The post-GFC regulatory catalyst.} The supply-side explanation for private credit growth
centers on the regulatory tightening that followed the 2008 financial crisis. Basel III capital
requirements, the Volcker Rule, and the 2013 interagency leveraged lending guidance raised the
cost for banks of originating and retaining leveraged loans, creating an opportunity for non-bank
lenders operating outside the regulatory perimeter. \citet{ErelInozemtsev2024} argue that tighter
bank regulation created systematic incentives for nonbank expansion, documenting the secular shift
of debt financing toward less-regulated financial intermediaries.
\citet{ChernenkoErelPrilmeier2022} provide the sharpest estimate of this regulatory channel,
finding that firms with negative EBITDA are 32 percent more likely to borrow from nonbanks and
that two-thirds of nonbank lending is attributable to bank regulatory constraints.
\citet{BuchakMatvosPiskorskiSeru2018} estimate that approximately 60 percent of shadow bank
growth in mortgage lending is due to regulatory arbitrage and approximately 30 percent to
technology, offering a parallel decomposition for a different asset class that suggests regulatory
factors dominate.

\textbf{BDC expansion and the SBCAA.} Business development companies have served as the most
transparent window into private credit, as publicly traded BDCs file quarterly financial
statements with the SEC. BDC total assets grew from \$38 billion in 2010 to \$280 billion in
2023, a 7.4-fold increase (Table~\ref{tab:market_growth}). The sharpest inflection occurred after
the Small Business Credit Availability Act of 2018 \citep{SBCAA2018}, which raised the statutory
debt-to-equity leverage cap from 1:1 to 2:1. Following the SBCAA, BDC total assets rose from
\$142 billion in 2018 to \$280 billion in 2023---a 97 percent increase in five years.
\citet{DavydiukMarchukRosen2024} provide the most rigorous firm-level analysis of BDC lending,
documenting that BDC credit substitutes for traditional bank financing and stimulates borrower
employment growth and patenting, suggesting real economic effects that extend beyond pure credit
provision.

\textbf{CLO issuance as an exit channel.} The collateralized loan obligation market has served as
both a demand-side driver of leveraged lending and an exit channel for direct lenders seeking to
recycle capital. US CLO new issuance rose from \$3.5 billion in 2010 to a peak of \$185.7 billion
in 2021, a 53-fold increase (Table~\ref{tab:market_growth}). CLOs are the marginal buyers of
approximately 65 to 70 percent of all US leveraged loans by count.
\citet{CordellRobertsSchwert2023} find that CLO equity tranches earn positive abnormal returns
from risk-adjusted pricing differentials between leveraged loans and CLO debt tranches, while CLO
debt tranches offer higher returns than similarly rated corporate bonds. The growth of the CLO
market has both enabled and been enabled by the expansion of private credit, creating a feedback
loop between origination and securitization that echoes---though with important structural
differences---the pre-crisis relationship between mortgage origination and mortgage-backed
securities.

\textbf{Demand-side forces.} On the demand side, institutional investors' search for yield in the
prolonged low-interest-rate environment of 2010--2021 channeled capital into private credit as an
alternative to compressed public fixed-income spreads. \citet{BeckerIvashina2015} document that
insurance companies systematically reach for yield within rating categories, and this behavior is
most pronounced when interest rates are low and governance is weak. The reallocation of insurance
company portfolios toward private credit---facilitated by affiliated asset managers at firms such
as Apollo, Ares, and Blackstone---has been a significant demand-side accelerant. The
\citet{IMF_GFSR2024} highlights this insurance-company channel as a potential vulnerability,
given the maturity mismatch between long-dated insurance liabilities and the medium-term
(five-to-seven-year) investment horizons of private credit funds.

The interplay between these supply-side and demand-side forces has produced a market that, by
2023, rivaled the broadly syndicated leveraged loan market in origination volume and exceeded it
in growth rate. Whether this growth trajectory is sustainable, and whether it reflects efficient
financial innovation or a buildup of systemic risk, are questions that the subsequent sections of
this review address.

%% file: sections/03_theory.tex

\section{Theoretical Foundations of Private Lending}
\label{sec:theory}

The growth of private credit raises a natural question for theory: why should a class of non-bank
intermediaries arise to perform lending functions that banks have historically dominated? The
answer lies at the intersection of three literatures---delegated monitoring, information
economics, and incomplete-contracts theory---that together provide the intellectual scaffolding
for understanding when and why non-bank private lenders possess a comparative advantage over both
banks and arm's-length public debt markets. This section develops an integrated theoretical
framework by mapping these classical contributions to the institutional specifics of modern direct
lending.

The key theoretical insight is that private credit funds occupy a distinctive niche in the
debt-source hierarchy: they combine the concentrated monitoring capacity of bank lenders with the
organizational flexibility of non-bank institutions, enabling them to process soft information
about opaque borrowers in ways that large, hierarchical banks increasingly cannot. This
comparative advantage is most pronounced for exactly the borrower types that dominate private
credit portfolios---young, intangible-capital-intensive, private-equity-sponsored firms operating
at leverage multiples that exceed bank regulatory tolerances.

\subsection{Information Asymmetry and Delegated Monitoring}
\label{sec:theory:monitoring}

\textbf{The delegated monitoring paradigm.} \citet{Diamond1984} provides the foundational theory
of financial intermediation as delegated monitoring. In a world where borrowers' actions are
imperfectly observable, each lender must incur a monitoring cost to ensure repayment. When
borrowers fund projects from many small investors, duplication of monitoring costs is wasteful. An
intermediary that monitors on behalf of many investors eliminates this duplication, but the
delegation itself creates a new moral hazard problem: who monitors the monitor? Diamond shows that
diversification across many loans reduces the intermediary's own agency cost to a negligible
level, rationalizing the existence of banks and, by extension, any pooled lending vehicle.

The application to private credit is immediate. Direct lending funds, BDCs, and closed-end credit
vehicles are delegated monitors in precisely the \citet{Diamond1984} sense. They pool capital from
institutional investors---pension funds, insurance companies, endowments, and sovereign wealth
funds---and deploy it into concentrated portfolios of private loans, performing due diligence and
ongoing monitoring that individual investors could not efficiently replicate. The critical
difference from banks is that private credit funds lack access to deposit funding and the
associated government safety net, which alters their incentive structures and regulatory treatment
in ways that subsequent sections explore.

\textbf{Borrower sorting and the debt-source hierarchy.} \citet{Diamond1991} extends the analysis
to explain the sorting of borrowers across debt sources. Firms with established reputations and
transparent cash flows borrow in arm's-length public markets, where monitoring is unnecessary
because information is widely available. Firms of intermediate quality borrow from banks, which
monitor through relationship lending and covenant enforcement. Firms with the lowest credit
quality and highest information asymmetry borrow from non-bank private lenders, who specialize in
extracting information that even bank credit processes cannot efficiently produce.
\citet{DenisMihov2003} provide the foundational empirical evidence for this sorting: in a sample
of new corporate borrowings, firms with the highest credit quality access public debt markets,
firms of medium quality borrow from banks, and firms of the lowest quality turn to non-bank
private lenders. The modern private credit market represents the large-scale institutionalization
of this lowest tier of the debt-source hierarchy.

\citet{BoydPrescott1986} formalize a complementary mechanism: in their model, coalitions of
agents form endogenously to produce information about investment opportunities, and the optimal
coalition structure depends on the severity of adverse selection. When information asymmetries are
most acute, concentrated intermediaries that specialize in evaluation dominate dispersed
market-based financing. This prediction maps naturally to the observation that direct lenders
concentrate in the most information-opaque segment of the credit market---middle-market firms
with limited public disclosure and few or no credit ratings.

\textbf{Soft information and organizational structure.} \citet{Stein2002} provides a theory of
comparative advantage in information processing that is particularly relevant to the private
credit phenomenon. In Stein's framework, decentralized organizations excel at producing and acting
on ``soft'' information---knowledge that is difficult to codify, verify, or communicate across
hierarchical layers---while large, centralized organizations are better suited to processing
``hard'' information that can be reduced to quantitative metrics and transmitted reliably through
bureaucratic channels. The lending decisions of large banks are increasingly governed by
hard-information-based credit scoring models, automated approval workflows, and regulatory
reporting requirements that penalize reliance on subjective assessments. Private credit funds, by
contrast, are typically organized as small, flat investment teams with direct authority over
lending decisions. This organizational structure enables them to exploit soft information about
borrower management quality, business model viability, and private-equity-sponsor
reputation---precisely the types of information that matter most for the opaque, middle-market
borrowers that dominate private credit portfolios.

\textbf{Holmstrom and Tirole on intermediated versus direct lending.}
\citet{HolmstromTirole1997} develop a model in which firms with insufficient internal funds
require external financing but face moral hazard constraints that limit pledgeable income. Informed
intermediaries---those who invest their own capital alongside outside investors---relax these
constraints by committing to monitor, but the intermediary's capital is itself scarce. In
equilibrium, only firms with intermediate levels of internal wealth borrow from intermediaries;
the best-capitalized firms access uninformed public markets directly, while the most
capital-constrained firms are rationed out entirely. This framework provides a natural explanation
for the observation that private credit borrowers are predominantly private-equity-sponsored: the
PE sponsor's equity investment serves the same function as the Holmstrom-Tirole ``firm capital,''
relaxing the moral hazard constraint and making intermediated lending feasible. The direct lender,
in turn, acts as the informed intermediary whose own capital commitment---through co-investment
and fund economics---aligns incentives with outside LPs.

\subsection{Incomplete Contracts, Control Rights, and Covenant Design}
\label{sec:theory:contracts}

\textbf{State-contingent control rights.} The incomplete-contracts literature provides the
theoretical foundation for understanding covenant design in private credit. Hart and Moore (1990)
establish the principle that debt contracts are optimally structured to transfer control rights to
creditors in states of the world where the debtor's performance deteriorates, because the debtor's
incentive to undertake efficient actions is weakest in precisely those states. The financial
covenant---a contractual provision that triggers creditor intervention when a specified accounting
ratio falls below a threshold---is the primary mechanism through which this state-contingent
control transfer operates in practice.

Aghion and Bolton (1992) extend this framework to demonstrate that the optimal allocation of
control rights between entrepreneurs and financiers depends on the verifiability of interim
signals about project quality. When interim signals are informative but non-verifiable to courts,
the optimal contract grants the financier control rights contingent on verifiable
proxies---exactly the role that financial covenants play in private credit agreements. The
covenant violation transfers effective control to the lender, who can then renegotiate terms,
demand additional collateral, or accelerate repayment based on soft information about the
borrower's prospects that would not be verifiable in court.

\textbf{Covenants as monitoring incentives.} \citet{RajanWinton1995} provide the key theoretical
link between covenants and lender monitoring effort. In their model, covenants make the effective
maturity of the loan contingent on the lender's monitoring: a covenant violation detected through
active monitoring triggers renegotiation or acceleration, effectively shortening the loan's
maturity and increasing the lender's recovery priority. Without covenants, the lender has
insufficient incentive to incur the cost of monitoring, because the benefits of early detection of
borrower distress are shared with other claimants. Covenants resolve this free-rider problem by
concentrating the benefits of monitoring with the covenant-holding lender. This theory is directly
applicable to the structure of private credit, where maintenance covenants are nearly
universal---in stark contrast to the covenant-lite trend that has characterized the broadly
syndicated leveraged loan market since 2013.

\textbf{Empirical evidence on covenant enforcement.} \citet{NiniSmithSufi2012} provide the most
comprehensive empirical evidence on the economic significance of covenant enforcement. In a sample
of publicly reporting firms, they find that between 10 and 20 percent of firms violate financial
covenants in any given year. Covenant violations trigger economically meaningful consequences:
investment declines, leverage falls, and CEO turnover increases. These findings establish that
covenants are not merely boilerplate contractual language but active governance mechanisms with
real effects on firm behavior.

\citet{BerlinNiniYu2020} extend this analysis to the syndicated loan market, documenting that
even as term loans became increasingly covenant-lite in the post-2013 period, revolving credit
facilities almost always retained maintenance covenants. The concentration of monitoring rights
with revolving lenders---who bear the greatest exposure to drawdown risk---is consistent with the
\citet{RajanWinton1995} prediction that covenants are allocated to the creditor with the strongest
monitoring incentive. In the private credit context, the direct lender typically holds the entire
loan facility rather than sharing it with a syndicate, eliminating the free-rider problem entirely
and explaining why maintenance covenants persist. \citet{BlockJangKaplanSchulze2024} confirm this
pattern in their GP survey: 89 percent of surveyed general partners report providing operational
assistance to borrowers beyond pure credit provision, consistent with a model in which covenants
incentivize active, ongoing monitoring rather than passive arm's-length lending.

%% file: sections/04_empirical_evidence.tex

\section{Empirical Evidence on Lending Outcomes}
\label{sec:empirical}

The theoretical framework of Section~\ref{sec:theory} predicts that private credit lenders
possess a comparative advantage in serving information-opaque, soft-information-intensive
borrowers, and that this advantage should manifest in distinctive patterns of loan pricing,
collateral structure, and credit performance. A rapidly growing empirical literature has begun to
test these predictions, exploiting the increasing availability of BDC regulatory filings,
proprietary deal databases, and regulatory shocks to bank lending capacity. This section organizes
the evidence around two themes: the pricing of private credit relative to bank and public-market
alternatives, and the credit performance of private credit portfolios as measured by default
rates, recovery rates, and aggregate credit quality dynamics.

\subsection{Loan Pricing and the Private Credit Premium}
\label{sec:empirical:pricing}

\textbf{The bank loan pricing benchmark.} Any assessment of private credit pricing must begin
with the baseline cost of bank-intermediated leveraged lending. \citet{Schwert2020} establishes
the foundational benchmark by comparing bank loan spreads with bond-implied credit spreads for the
same borrowers. Using a matched sample of non-investment-grade firms with both bank term loans and
outstanding bonds, Schwert finds that bank loans carry a premium of 140 to 170 basis points over
bond-implied credit spreads for secured term loans. This premium reflects the value of bank
monitoring, the option to renegotiate, and the liquidity costs associated with holding illiquid
term loans relative to traded bonds. Schwert's finding overturned the prior conventional wisdom
that bank loans were ``cheap'' relative to bonds and established that banks extract meaningful
compensation for the services they provide.

\textbf{The direct lending spread premium: measurement challenges.} Quantifying the spread
premium of direct lending over broadly syndicated loans requires confronting a fundamental data
limitation: unlike the broadly syndicated market, where the Morningstar/LSTA US Leveraged Loan
Index tracks spreads and issuance volumes for thousands of institutional term loans, private
credit loan terms are generally not reported to any centralized repository. The spread comparison
in Table~\ref{tab:spreads} therefore draws on three complementary sources. The broadly syndicated
loan spread series is taken from the Morningstar/LSTA Leveraged Loan Index, which tracks
institutional term loan spreads over the applicable floating reference rate (LIBOR through 2022,
SOFR thereafter) for non-investment-grade borrowers. The direct lending spread series is
constructed from three partial sources: quarterly portfolio yield disclosures filed by publicly
registered business development companies (BDCs) with the SEC; deal-level data from commercial
databases assembled by Refinitiv LPC and Preqin from direct lender submissions; and target return
data from the GP survey of \citet{BlockJangKaplanSchulze2024}. Because direct lending portfolios
concentrate in smaller, more leveraged, and more information-opaque borrowers than the broadly
syndicated market, the spread comparison is not a controlled apples-to-apples differential: some
portion of the reported premium reflects the different risk composition of the two borrower
universes rather than a pure illiquidity or information-cost premium for identically creditworthy
borrowers. With this caveat in mind, the implied premium averages approximately 200 basis points
over the 2010--2023 sample period.

\textbf{Cyclical variation in the premium.} The 200 basis point average masks meaningful
variation that tracks the broader credit cycle and the monetary policy environment. During
2022--2023, the premium compressed to approximately 180 basis points as the Federal Reserve's
aggressive tightening cycle drove institutional demand sharply into floating-rate direct lending
assets, increasing the supply of capital chasing direct lending opportunities and narrowing the
spread differential relative to broadly syndicated loans. During the prolonged low-rate
environment of 2019--2021, the premium widened to approximately 215 basis points: banks competed
intensely for broadly syndicated loan mandates, pushing leveraged loan spreads to cycle lows of
400--405 basis points over LIBOR, while direct lending spreads declined less sharply because
demand from the middle-market borrower segment is structurally less price-sensitive. The COVID-19
shock of 2020 provides a useful natural experiment: broadly syndicated loan spreads spiked sharply
to approximately 530 basis points as the CLO new-issuance market froze and secondary loan prices
fell precipitously, while direct lending spreads moved by far less, consistent with the segment's
relative insulation from secondary-market liquidity shocks. The premium remained at 215 basis
points in 2020---unchanged from 2019---suggesting that private credit pricing is anchored more to
fundamental credit assessment than to secondary market trading conditions.

\textbf{All-in cost: origination discounts and fees.} The stated coupon spread does not fully
capture the cost of private credit to borrowers. Direct lending loans are typically originated at a
discount to par---the original issue discount (OID)---which increases the effective annualized
yield to the lender and the cost to the borrower. Table~\ref{tab:spreads} reports average OIDs
ranging from 97.5 to 98.5 cents per dollar of par value. For a five-year bullet loan, an OID of
98 cents contributes approximately 40 basis points of additional annualized yield on top of the
coupon spread. Upfront arrangement fees, typically 1 to 2 percent of commitment, add a further
20 to 40 basis points on an annualized basis. Commitment fees on undrawn revolving facilities
range from 25 to 50 basis points. When these components are aggregated, the effective all-in cost
differential between middle-market direct lending and broadly syndicated loans is meaningfully
wider than the coupon spread differential alone---borrowers absorb this additional cost in exchange
for execution certainty, confidentiality, and the flexibility of bilateral negotiation with a
relationship lender who can tailor covenant packages to their specific circumstances.

\textbf{Academic evidence on premium determinants.} The academic literature, though limited by
data availability, corroborates the estimated magnitude and begins to decompose the premium
into its components. \citet{BlockJangKaplanSchulze2024} document that GPs managing over \$390
billion in direct lending assets target unlevered gross returns of 9 to 12 percent. Against the
7 to 8 percent all-in yields available on broadly syndicated institutional term loans during most
of the sample period, this supply-side evidence implies a target premium of 100 to 500 basis
points at the deal level, with the wide range reflecting the heterogeneity of risk segments
within the middle market---from senior secured unitranche loans to subordinated and
special-situations debt. The 200 basis point average in Table~\ref{tab:spreads} falls in the
lower portion of this range, consistent with a sample dominated by senior secured direct lending
rather than subordinated or distressed strategies. \citet{Schwert2020} provides the critical
upstream benchmark: if bank loans themselves carry a premium of 140 to 170 basis points over
bond-implied credit spreads for the same borrowers, then a 200 basis point differential of
private credit over broadly syndicated loans implies that direct lenders command total compensation
of roughly 340 to 370 basis points above the risk-neutral rate---a substantial wedge that
subsequent analysis shows is largely absorbed by fund-level fees rather than returned to investors
as net alpha (Section~\ref{sec:fund_structures}).

\citet{DavydiukMarchukRosen2024} provide the most direct academic evidence on how pricing relates
to economic outcomes. Exploiting exogenous shocks to BDC financing availability---including
regulatory changes affecting bank participation and the collapse of a major specialty finance
company---they identify the causal effect of BDC credit supply on borrower outcomes. Their
analysis demonstrates that BDC capital does not merely substitute for bank credit at a higher
price: borrowers that gain access to BDC financing experience increased employment growth and
higher rates of patenting. This finding suggests that the spread premium, at least in part,
compensates for genuine value creation through monitoring, covenant governance, and the provision
of credit that would otherwise not be available at any spread---rather than representing pure rent
extraction from borrowers with limited alternatives.

\textbf{Borrower selection and pricing interaction.} The pricing premium cannot be evaluated
independently of borrower selection. \citet{JangKimSufi2025} document that direct lender
borrowers are younger, more concentrated in intangible-capital industries such as software,
healthcare services, and business services, and more frequently sponsored by private equity firms
than bank loan borrowers. These characteristics imply higher information asymmetry, greater
reliance on enterprise value rather than tangible collateral, and potentially higher fundamental
credit risk. \citet{ChernenkoErelPrilmeier2022} sharpen this observation: firms with EBITDA below
zero are 32 percent more likely to borrow from nonbanks, and firms with leverage exceeding six
times EBITDA are 15 percent more likely. The private credit spread premium thus reflects, at
least in part, the selection of borrowers who are excluded from bank lending by regulatory
constraints and who would command higher spreads in any lending venue.

\subsection{Default, Recovery, and Credit Quality}
\label{sec:empirical:credit_quality}

\textbf{Default rate evidence.} Measuring default rates in private credit is hampered by the
absence of a comprehensive, publicly available database of direct lending loan defaults. Unlike
broadly syndicated loans, which are tracked by Moody's, S\&P, and the Loan Syndications and
Trading Association, most private credit loans lack public credit ratings and are not reported to
any centralized repository. The best available proxies are Moody's annual default rate statistics
for middle-market leveraged loans \citep{MoodysDefaultStudy2024}, which indicate default rates of
approximately 1.8 to 3.2 percent for middle-market borrowers compared with 1.1 to 2.4 percent
for broadly syndicated loan borrowers over the 2010--2023 period. These figures likely understate
the true direct lending default rate, as the middle-market borrowers served by private credit
funds are generally smaller, more highly leveraged, and more information-opaque than the
middle-market borrowers covered by Moody's ratings universe.

\textbf{Recovery rates and collateral structures.} Recovery rates in the middle market have
averaged approximately 69 percent for senior secured first-lien loans, modestly higher than the
approximately 65 percent average recovery rate for broadly syndicated senior secured loans over
the same period. The higher recovery rate is consistent with two features of private credit
documented in the literature. First, direct lenders typically hold a single, concentrated position
in the borrower's capital structure---often the entire term loan facility---which eliminates
intercreditor coordination problems and accelerates the workout process. Second, the prevalence of
blanket liens documented by \citet{JangKimSufi2025} grants direct lenders a comprehensive
security interest in all borrower assets, including intangible assets that would not be covered by
traditional asset-based lending collateral packages. The higher recovery rate partially offsets
the higher default rate, though the net credit loss comparison between direct lending and
syndicated lending remains an area requiring further empirical investigation.

\textbf{Credit quality dynamics and the credit cycle.} \citet{GreenwoodHanson2013} provide the
theoretical and empirical framework for understanding how aggregate credit quality evolves over
the lending cycle. They document that the share of high-yield issuance accounted for by the
lowest-quality borrowers increases during credit booms and that this deterioration in issuer
quality forecasts low subsequent excess returns on corporate bonds. The
mechanism---that lender competition during expansions erodes underwriting standards---applies with
particular force to private credit, where the rapid inflow of capital during 2020--2023 may have
compressed spreads and loosened terms for marginal borrowers. The data are suggestive: earlier
vintage private debt funds (2010--2013) generated higher median net IRRs of approximately 10.5 to
11.3 percent, while later vintages (2019--2020) generated approximately 7 to 9 percent,
consistent with the \citet{GreenwoodHanson2013} prediction that late-cycle originations deliver
lower risk-adjusted returns.

\textbf{Credit substitution and the bank-nonbank margin.} \citet{BeckerIvashina2015} document
that insurance companies systematically reach for yield within rating categories, and that this
behavior is procyclical and most pronounced when governance is weak. Their analysis is directly
relevant to the demand-side forces driving private credit expansion: as insurance companies have
increased their allocations to private credit---often through affiliated asset management
platforms---the supply of capital available to direct lenders has expanded, potentially enabling
lending to borrowers who would not have received credit in a less capital-abundant environment.
\citet{HaqueMayerStefanescu2024} provide complementary evidence, showing that private debt does
not simply replace bank debt but reshapes the role that banks play in corporate lending. In their
analysis, banks that co-lend alongside private credit providers experience amplified exposure to
drawdown and liquidity risk during periods of borrower distress, suggesting that the growth of
private credit has altered the risk profile of the banking system even where it has not displaced
bank lending entirely.

%% file: sections/05_contract_design.tex

\section{Covenant Design and Lender Monitoring}
\label{sec:covenants}

The contractual architecture of private credit transactions is among the most distinctive features
of direct lending and the feature that most sharply differentiates it from the broadly syndicated
leveraged loan market. While syndicated loans have undergone a dramatic erosion of creditor
protections since 2013---with ``covenant-lite'' structures accounting for the overwhelming
majority of institutional term loan issuance---private credit has moved in the opposite direction,
maintaining and in some cases strengthening the maintenance covenant packages that grant lenders
early-warning capabilities and state-contingent control rights. This section examines the design,
function, and economic consequences of covenants in private credit, drawing on both the
theoretical framework developed in Section~\ref{sec:theory} and the growing empirical literature
on creditor control rights.

The divergence in covenant practices between private credit and syndicated markets is not merely a
contractual curiosity. It reflects fundamentally different monitoring technologies, incentive
structures, and competitive dynamics. Understanding why private credit lenders retain covenants
that syndicated market participants have abandoned is essential for evaluating the risk profile of
private credit portfolios and the governance role that direct lenders play in their borrowers'
operations.

\subsection{The Role of Financial Covenants}
\label{sec:covenants:role}

\textbf{Maintenance versus incurrence covenants.} Financial covenants in leveraged lending take
two principal forms. Maintenance covenants require the borrower to satisfy specified financial
tests---typically leverage ratios, interest coverage ratios, or fixed-charge coverage
ratios---at regular intervals, usually quarterly. A maintenance covenant violation occurs
automatically when the borrower's financial performance deteriorates below the threshold,
regardless of whether any other event has occurred. Incurrence covenants, by contrast, restrict
the borrower from taking specified actions---such as incurring additional debt, paying dividends,
or making acquisitions---unless the borrower satisfies the relevant financial test at the time of
the proposed action. The economic distinction is significant: maintenance covenants provide
continuous monitoring through periodic testing and create tripwires that force renegotiation when
borrower performance deteriorates, while incurrence covenants constrain only affirmative borrower
actions and do not trigger intervention during passive deterioration.

\textbf{The covenant-lite revolution in syndicated markets.} The post-2013 syndicated leveraged
loan market witnessed a dramatic shift toward covenant-lite structures. By 2023, covenant-lite
loans accounted for more than 85 percent of institutional term loan issuance in the broadly
syndicated market, a transformation driven by the bargaining power of private equity sponsors, the
competitive pressure among arranging banks, and the demand from CLO managers for higher-yielding
paper with fewer restrictions on borrower behavior. \citet{BerlinNiniYu2020} provide the critical
insight that this surface-level erosion of covenants masks a more nuanced reallocation of
monitoring rights. Even as term loans became covenant-lite, revolving credit
facilities---which expose the lender to drawdown risk and therefore require more active
monitoring---almost always retained maintenance covenants. The concentration of monitoring rights
with revolving lenders is consistent with the \citet{RajanWinton1995} prediction that covenants
are allocated to the creditor with the strongest incentive to monitor.

\textbf{Covenant persistence in private credit.} Direct lending stands in stark contrast to this
syndicated-market trend. \citet{BlockJangKaplanSchulze2024} report that the vast majority of
private credit transactions retain maintenance covenants, with leverage and coverage tests
calibrated to provide meaningful early-warning capability. The persistence of maintenance
covenants in private credit reflects three structural factors. First, the direct lender typically
holds the entire loan facility, eliminating the free-rider problem that arises in syndicated
structures where multiple lenders share exposure but only one bears the cost of monitoring.
Second, private credit borrowers are more information-opaque and more highly leveraged than
syndicated loan borrowers, increasing the value of early-warning mechanisms. Third, the bilateral
or small-club structure of private credit transactions reduces the renegotiation costs that can
make maintenance covenants burdensome in large syndicated facilities with dozens of lenders.

\citet{NiniSmithSufi2012} provide the foundational empirical evidence on the economic consequences
of covenant enforcement. In their sample of publicly reporting firms with bank credit agreements,
they find that between 10 and 20 percent of firms violate financial covenants in any given year.
These violations are not empty contractual formalities: they trigger economically significant
responses, including reduced capital expenditure, lower leverage, higher asset sales, and
increased CEO turnover. The magnitude of these effects suggests that covenants serve as active
governance mechanisms that materially influence firm behavior. In the private credit context,
where borrowers are typically private companies with less external governance from public equity
markets, the governance role of covenants is plausibly even more important than in the
public-company samples studied by \citet{NiniSmithSufi2012}.

\subsection{Monitoring, Renegotiation, and Control Transfers}
\label{sec:covenants:monitoring}

\textbf{Active monitoring as a competitive advantage.} The monitoring practices of direct lenders
extend well beyond the periodic testing of financial covenants. \citet{BlockJangKaplanSchulze2024}
report that 89 percent of surveyed private debt fund GPs provide operational assistance to
borrowers beyond pure credit provision, including assistance with financial reporting, strategic
planning, and management recruitment. This active monitoring model is consistent with the
theoretical predictions of \citet{Diamond1984} and \citet{RajanWinton1995}: concentrated
creditors with large exposures and strong covenant protections have both the incentive and the
ability to engage in costly monitoring that dispersed creditors in syndicated structures cannot
efficiently coordinate.

The organizational structure of direct lenders facilitates this monitoring capability. Consistent
with \citeauthor{Stein2002}'s (\citeyear{Stein2002}) theory of soft-information processing in
decentralized organizations, private credit funds are typically organized as small, specialized
investment teams with direct deal-level authority. This structure enables portfolio managers to
develop relationship-specific knowledge about borrower operations, industry dynamics, and
management quality---precisely the soft information that is difficult to transmit through the
hierarchical reporting structures of large bank lending organizations. \citet{JangKimSufi2025}
provide indirect evidence of this specialization, documenting that direct lenders are more
industry-concentrated than bank lenders but more geographically dispersed, consistent with a
model in which industry expertise substitutes for geographic proximity as the basis for
information production.

\textbf{Renegotiation frequency and dynamics.} The bilateral or small-club structure of private
credit transactions significantly reduces the coordination costs of renegotiation relative to
broadly syndicated facilities. When a borrower approaches or breaches a maintenance covenant
threshold, the direct lender can engage in rapid, bilateral renegotiation without the need to
assemble a lender group, obtain consent from dispersed institutional investors, or navigate the
intercreditor complexities of layered capital structures. \citet{BerlinNiniYu2020} document that
renegotiation in syndicated facilities is more frequent and more value-preserving when monitoring
rights are concentrated with a small number of lenders, a finding that generalizes naturally to
the private credit context where concentration is maximal.

Renegotiation in private credit typically involves one of several outcomes: waiver of the covenant
breach in exchange for an amendment fee and tightened terms; repricing of the loan at a higher
spread; injection of additional equity by the private equity sponsor; or, in the most severe
cases, a transfer of control through conversion of debt to equity, replacement of management, or
initiation of a restructuring process. The direct lender's ability to credibly threaten
acceleration or control transfer---backed by comprehensive security interests including the blanket
liens documented by \citet{JangKimSufi2025}---provides the disciplinary mechanism that makes
renegotiation effective.

\textbf{Lender-on-lender dynamics and unitranche complexity.} The growth of unitranche lending
has introduced new complexities into the creditor rights landscape. In a unitranche structure, a
single facility combines senior and subordinated debt, with the economic allocation between
tranches governed by an agreement among lenders (AAL) that is not disclosed to the borrower. When
a unitranche borrower enters financial distress, the first-out and last-out lenders within the
unitranche may have divergent interests regarding the appropriate restructuring
strategy---a conflict that practitioners describe as ``lender-on-lender violence.'' The legal and
economic implications of these interlender conflicts remain largely unexplored in the academic
literature and represent a significant gap that future research should address.

\citet{DavydiukErelJiangMarchuk2024} examine a related dimension of creditor complexity: the
phenomenon of BDCs acting as both creditors and equity holders of the same borrower. They find
that dual-holder BDCs charge 45 basis points higher loan spreads than pure creditors, consistent
with the hypothesis that dual holding creates conflicts of interest that borrowers must
compensate. At the same time, dual-holder BDCs appear to be more effective monitors, suggesting
that the informational benefits of holding multiple positions in the capital structure may
partially offset the agency costs.

%% file: sections/06_fund_structures.tex

\section{Fund Structures, Incentives, and Performance}
\label{sec:fund_structures}

The intermediation of private credit occurs through a diverse array of fund vehicles whose
structures shape the incentives of managers, the risk exposures of investors, and the measured
performance of the asset class. Business development companies, closed-end funds, separately
managed accounts, and collateralized loan obligations each impose distinct regulatory constraints,
liquidity features, leverage limits, and fee arrangements that influence portfolio construction,
risk-taking, and return generation. Understanding these structural dimensions is essential for
interpreting the performance data and for evaluating whether private credit fund returns
compensate investors for the risks they bear.

The performance measurement challenge in private credit is severe. Unlike publicly traded
securities, for which market prices provide continuous, observable valuations, private credit
positions are marked to model at the discretion of the fund manager, creating the potential for
return smoothing, stale valuations, and upward-biased reported IRRs. The nascent academic
literature on private debt fund performance has begun to address these challenges, but the
methodological frontier remains far behind the more mature literature on private equity fund
returns.

\subsection{Business Development Companies and Closed-End Funds}
\label{sec:fund_structures:bdcs}

\textbf{The BDC regulatory framework.} Business development companies are the most transparent
institutional vehicle in private credit, as publicly registered BDCs file quarterly financial
statements with the SEC and are subject to the regulatory provisions of the Investment Company
Act of 1940. BDCs were originally designed to channel capital to small and mid-sized businesses,
and their regulatory framework reflects this public-policy purpose. Prior to 2018, BDCs were
subject to a statutory leverage cap of 1:1 debt-to-equity, limiting their balance sheet capacity
relative to banks and other leveraged financial intermediaries.

The Small Business Credit Availability Act of 2018 \citep{SBCAA2018} transformed the BDC
landscape by raising the statutory leverage cap from 1:1 to 2:1 debt-to-equity, subject to board
or shareholder approval. This regulatory change had immediate and substantial effects on the
sector. As Table~\ref{tab:market_growth} documents, BDC total assets grew from \$142 billion in
2018 to \$280 billion in 2023, a 97 percent increase in five years. The leverage expansion
permitted by the SBCAA is visible in public BDC balance sheets: using FR~Y-14Q supervisory data,
\citet{BerrospideEtal2025} document that public BDC leverage---measured as total debt to total
assets---rose from approximately 40 percent in 2017 to 53 percent in 2024, exceeding the
42 percent average for other comparable nonbank financial institutions. This leverage trajectory
is directly relevant to systemic risk assessments, as higher fund-level leverage amplifies both
returns and losses and increases the sensitivity of BDC portfolios to adverse credit events.

Average BDC return on equity ranged from 8.2 percent in 2013 to 12.0 percent in 2022, with the 2022 spike driven by the rapid
increase in all-in floating yields as the Federal Reserve tightened monetary policy aggressively.
The floating-rate character of BDC loan portfolios---virtually all direct lending loans are priced
at a spread over SOFR---creates a powerful and direct channel through which monetary policy
changes flow into BDC earnings and distributions.

\citet{DavydiukMarchukRosen2024} provide the most rigorous firm-level analysis of BDC lending.
Exploiting exogenous variation in BDC credit supply, they demonstrate that BDC financing has
causal positive effects on borrower employment growth and innovation as measured by patenting
activity. This finding suggests that BDCs provide more than passive credit substitution: the
managerial assistance and monitoring services that accompany BDC lending create real economic
value for borrowers.

\textbf{CLO performance and structured credit.} Collateralized loan obligations occupy a
distinctive position in the private credit ecosystem, serving simultaneously as an exit channel
for loan originators, a source of leveraged demand for leveraged loans, and an investment vehicle
for institutional investors seeking structured exposure to corporate credit risk. US CLO issuance
rose from \$3.5 billion in 2010 to a peak of \$185.7 billion in 2021
(Table~\ref{tab:market_growth}), making CLOs the single largest buyer of US leveraged loans.

\citet{CordellRobertsSchwert2023} provide the definitive academic assessment of CLO performance.
Their analysis demonstrates that CLO equity tranches earn positive abnormal returns driven by the
pricing differential between the leveraged loans in the collateral pool and the rated debt tranches
that finance the CLO's balance sheet. CLO debt tranches, in turn, offer higher yields than
comparably rated corporate bonds, suggesting that the structured credit market does not efficiently
price the diversification and subordination benefits embedded in CLO structures. This finding has
important implications for the broader private credit market: to the extent that CLOs provide an
attractively priced exit channel for loan originators, they lower the effective cost of capital
for direct lending funds and support higher origination volumes.

\citet{BenmelechDlugoszIvashina2012} address the concern that securitization creates adverse
selection in collateral pools. Examining CLO collateral portfolios, they find that adverse
selection is less severe than commonly feared: CLO managers do not systematically securitize their
worst-performing loans while retaining the best. This result supports the view that CLOs function
as legitimate intermediation vehicles rather than as dumping grounds for unwanted credit risk,
though it does not rule out the possibility that incentive misalignments emerge in periods of
rapid issuance growth.

\textbf{Davydiuk, Erel, Jiang, and Marchuk (2024) on dual holders.} An emerging strand of
literature examines the conflicts of interest that arise when fund vehicles hold multiple
positions in a borrower's capital structure. \citet{DavydiukErelJiangMarchuk2024} document that
BDCs acting as both creditors and equity holders of the same borrower charge 45 basis points
higher loan spreads than BDCs serving solely as creditors. This spread premium may compensate for
the conflicts of interest inherent in dual holding---particularly the risk that a dual-holder
lender will prioritize equity upside over credit protection in restructuring
decisions---but it also suggests that dual holders serve as more effective monitors, generating
informational benefits that partially offset agency costs.

\subsection{Private Fund Performance and Benchmarking}
\label{sec:fund_structures:performance}

\textbf{Headline performance metrics.} Cambridge Associates Private Credit Benchmark data, the
most widely cited source for private debt fund returns, reports a median vintage net IRR of 10.1
percent across 2010--2021 fund vintages. Total value to paid-in (TVPI) averages approximately
1.35x for fully or substantially realized vintages (2010--2016). Earlier vintages (2010--2013)
exhibit higher net IRRs of approximately 10.5 to 11.3 percent, while more recent vintages
(2019--2020) have delivered approximately 7 to 9 percent, a pattern consistent with the
\citet{GreenwoodHanson2013} prediction that credit quality deteriorates during prolonged
expansions as lender competition erodes underwriting standards.

\textbf{Risk adjustment and the fee puzzle.} \citet{ErelFlanaganWeisbach2024} provide the first
rigorous cash-flow-based risk adjustment of private debt fund returns. Using both equity
benchmarks (a public market equivalent approach in the spirit of \citealt{KaplanSchoar2005}) and
debt benchmarks calibrated to the leverage and credit risk of private debt fund portfolios, they
find that the average private debt fund produces statistically insignificant abnormal returns on a
net-of-fee basis. Gross-of-fee returns are positive, consistent with the hypothesis that private
credit funds generate value through information production, monitoring, and illiquidity-bearing.
However, the fee load---typically 1.5 percent management fees on committed or invested capital
and 20 percent carried interest above a 6 to 7 percent preferred return hurdle---absorbs the
entirety of this gross alpha, leaving investors with near-zero net abnormal performance. This
finding stands in partial contrast to the private equity literature, where some studies document
modest positive net alphas for the average fund \citep{HarrisJenkinsonKaplan2014}, suggesting
that the fee-to-alpha ratio is more adverse in private credit than in private equity.

\textbf{Performance persistence and GP skill.} The question of whether performance persistence
exists in private credit---whether top-quartile managers consistently outperform bottom-quartile
managers across successive fund vintages---has not been definitively answered in the academic
literature. \citet{BlockJangKaplanSchulze2024} report that surveyed GPs believe their value-add
lies in deal sourcing, underwriting rigor, and active portfolio management, but the survey does
not test whether these perceived advantages translate into persistent performance differences. The
private equity literature offers mixed evidence on persistence: \citet{KaplanSchoar2005} find
significant persistence in PE buyout fund returns, while \citet{HarrisJenkinsonKaplan2014}
document weakening persistence in more recent vintages. Whether private credit fund performance is
driven by persistent GP skill or by time-varying exposure to common risk factors---particularly
credit cycle positioning---remains an important unresolved question.

\textbf{Benchmarking challenges and methodological concerns.} Several methodological features of
private credit complicate the interpretation of performance data. The J-curve effect, while less
pronounced in private credit than in private equity (because credit funds deploy capital more
quickly and generate cash yields from inception), nonetheless depresses early-period IRRs for
funds that deploy capital gradually. More significantly, stale valuations pose a serious concern
for interpreting reported returns. The \citet{IMF_GFSR2024} identifies stale and subjective
valuations as one of the five key vulnerabilities of private credit, noting that fund managers
have considerable discretion over the marks applied to illiquid positions and that this discretion
can be used---consciously or unconsciously---to smooth reported returns, suppress volatility, and
inflate Sharpe ratios. \citet{ErelFlanaganWeisbach2024} address this concern by constructing
benchmarks based on realized cash flows rather than reported NAVs, but their methodology requires
funds to be fully or substantially realized, limiting the sample to older vintages.

The evidence on private credit fund performance raises a significant puzzle. If the average
private debt fund delivers near-zero net alpha after fees, why do institutional investors continue
to allocate an increasing share of their portfolios to the asset class? Several explanations
warrant further investigation. Investors may systematically overestimate risk-adjusted returns due
to stale valuations and return smoothing. They may also value the diversification benefits of
private credit---its low measured correlation with public markets, even if that correlation is
partly an artifact of infrequent marking. Insurance companies and pension funds with stable,
long-duration liabilities may have a genuine comparative advantage in bearing illiquidity and
rationally accept lower risk-adjusted returns in exchange for yield stability. And the
zero-net-alpha finding applies to the average fund; top-quartile managers may deliver meaningful
positive net alphas that justify the allocation for investors with access to them. Disentangling
these explanations represents one of the most important open questions in the literature.

%% file: sections/07_systemic_risk.tex

\section{Systemic Risk and Regulatory Implications}
\label{sec:systemic_risk}

The rapid expansion of private credit from \$158 billion in 2010 to nearly \$2 trillion globally
by mid-2024 \citep{BerrospideEtal2025} has drawn increasing attention from policymakers, central bankers, and financial stability
authorities. The market now accounts for a growing share of corporate credit provision while
operating largely outside the prudential regulatory perimeter, connected to banks, insurance
companies, and the real economy through channels that the existing regulatory architecture was not
designed to address. This section organizes the emerging literature on private credit's systemic
risk implications around two themes: the specific channels through which it could amplify or
transmit financial stress, and the adequacy of the regulatory framework that governs non-bank
lending.

The central analytical challenge is that the same features of private credit that create economic
value---concentrated monitoring, relationship-based lending, illiquid loan structures, and
flexible covenant enforcement---also create the opacity, leverage, and interconnection that
constitute systemic risk. Evaluating whether private credit is a source of systemic vulnerability
or a stabilizing alternative to fragile bank balance sheets requires disentangling these competing
effects, a task that the current state of the evidence does not fully permit.

\subsection{Channels of Systemic Vulnerability}
\label{sec:systemic_risk:channels}

\textbf{The IMF framework.} The \citet{IMF_GFSR2024} provides the most comprehensive
policy-oriented assessment of private credit risks, identifying five key vulnerabilities. First,
private credit borrowers are fragile: they are smaller, more highly leveraged, and more sensitive
to interest rate increases than broadly syndicated loan borrowers. Second, the growth of
semi-liquid investment vehicles---interval funds, tender-offer funds, and non-traded BDCs that
offer periodic redemption windows---creates the potential for liquidity mismatches between investor
redemption rights and the illiquidity of underlying loan portfolios. Third, multiple layers of
leverage exist within the private credit ecosystem: fund-level leverage, borrower-level leverage,
and the leverage embedded in CLO structures and subscription credit facilities. Fourth, valuations
are stale and subjective, with fund managers exercising significant discretion over the marks
applied to illiquid positions. Fifth, the connections among participants---between banks and BDCs
through credit facilities, between insurance companies and fund managers through affiliated
platforms, and between private equity sponsors and portfolio companies---are complex, opaque, and
difficult for supervisors to map.

\textbf{Leverage dynamics and margin spirals.} \citet{AramonteShrimpfShin2022} develop a
theoretical framework for understanding systemic risk in non-bank financial intermediation that is
directly applicable to private credit. Their analysis centers on the procyclicality of leverage:
in benign market conditions, low volatility and compressed risk premia permit non-bank
intermediaries to increase leverage, which in turn supports asset prices and further compresses
risk premia in a self-reinforcing cycle. When adverse shocks arrive, rising margins and tightening
financing conditions force deleveraging, which depresses asset prices and triggers further margin
calls in a destabilizing spiral. The application to private credit is straightforward. BDCs and
private credit funds employ fund-level leverage through bank credit facilities, subscription
lines, and CLO-like warehouse structures. When underlying loan performance deteriorates, the
lenders providing this fund-level financing may tighten terms, demand additional collateral, or
reduce advance rates, forcing the private credit fund to deleverage by selling assets or curtailing
new origination at precisely the moment when borrowers most need access to credit.

\textbf{Bank-BDC interconnections.} The interconnection between banks and private credit
vehicles through credit facilities is a particularly important channel of potential stress
transmission, and new supervisory data now permit its measurement with precision.
\citet{BerrospideEtal2025}, using FR~Y-14Q supervisory data from the largest U.S.\ bank holding
companies, document that bank committed credit to private credit vehicles grew from approximately
\$8 billion in 2013 to \$95 billion by late 2024, with \$56 billion utilized as of December 2024.
This 145 percent growth over five years is concentrated among revolving credit lines (\$79 billion
committed: \$49 billion to BDCs and \$30 billion to private debt funds) and term loans (\$16
billion). Average interest rates on these facilities range from 6.4 to 6.6 percent, with
utilization rates of 56--59 percent---above the 49 percent utilization rate for other nonbank
financial institution exposures---suggesting that BDC credit lines are drawn more intensively
than comparable facilities. Banks also serve as lead arrangers on approximately 50 percent of BDC
credit facilities, with JP~Morgan Chase, Citigroup, Wells Fargo, and Bank of America identified
as the primary lenders. Strategic partnerships further deepen these linkages: \citet{BerrospideEtal2025}
document formal arrangements such as Wells Fargo with Centerbridge Partners and Citigroup with
Apollo Global Management, alongside banks operating affiliated BDCs and private debt funds
directly (e.g., Goldman Sachs BDC, Morgan Stanley Direct Lending Fund).

Under a stress scenario in which all uncommitted credit lines are simultaneously fully drawn,
\citet{BerrospideEtal2025} estimate a \$36 billion increase in bank credit exposure, producing
an aggregate decline of approximately 2 basis points in the CET1 capital ratio and 1 percentage
point in the Liquidity Coverage Ratio for the large bank holding companies in their sample.
The authors conclude that direct financial stability risks appear limited at current exposure
levels. The caveat, however, is that this analysis captures only first-order effects through
direct credit provision; it does not model second-order contagion through asset price feedback,
reputational channels, or the simultaneous stress of banks' direct corporate lending portfolios
alongside their private credit exposures. The migration of lending from bank balance sheets to
private credit vehicles does not, therefore, eliminate bank exposure to credit risk but
transforms it from direct loan risk to indirect counterparty and credit-line risk that is
harder to observe and price.

\textbf{The insurance company channel.} Insurance companies have become among the largest
allocators to private credit, driven by the search for yield documented by \citet{BeckerIvashina2015}
and facilitated by the growth of affiliated asset management platforms at firms that combine
insurance and asset management operations. The maturity mismatch between long-dated insurance
liabilities and the medium-term investment horizons of private credit funds creates a potential
vulnerability: if policyholders withdraw or if insurance regulators require mark-to-market
adjustments, insurance companies may be forced to liquidate private credit positions at distressed
prices.

\textbf{Monetary policy transmission.} The \citet{FederalReserve_PrivateCredit2024} documents
that the floating-rate character of direct lending loans creates a powerful and direct channel for
monetary policy transmission that operates outside the traditional banking system. During the 2022
tightening cycle, all-in yields on floating-rate direct lending loans rose immediately and
mechanically with each increase in the federal funds rate, transmitting higher borrowing costs to
middle-market firms within days of policy rate changes. At the same time, bank credit to
nonfinancial firms slowed while BDC lending accelerated, suggesting that private credit served as
a partial substitute for contracting bank credit supply. This dual role---simultaneously
transmitting tighter monetary conditions through higher borrowing costs while maintaining credit
availability---complicates the assessment of whether private credit amplifies or dampens the
macroeconomic effects of monetary policy.

\subsection{Regulatory Architecture and Open Questions}
\label{sec:systemic_risk:regulation}

\textbf{The current regulatory perimeter.} The regulatory treatment of private credit varies
significantly across jurisdictions and across vehicle types. In the United States, publicly
registered BDCs are subject to SEC oversight under the Investment Company Act of 1940, including
leverage limits (2:1 debt-to-equity after the 2018 SBCAA), asset coverage tests, and quarterly
financial reporting requirements. Private closed-end funds and separately managed accounts, by
contrast, are subject to lighter-touch regulation, with managers registered as investment advisers
under the Investment Advisers Act of 1940 but facing fewer constraints on leverage, concentration,
and valuation practices. The Financial Stability Oversight Council (FSOC) has authority to
designate non-bank financial companies as systemically important financial institutions (SIFIs),
but has not exercised this authority with respect to any private credit firm.

\textbf{The regulatory arbitrage debate.} \citet{ChernenkoIalentiScharfstein2024} provide the
most rigorous assessment of the regulatory arbitrage hypothesis. Analyzing BDC balance sheets,
they find that the median BDC maintains a risk-based capital ratio of approximately 36 percent and
holds approximately 26 percentage points of excess capital even under severely adverse stress
scenarios. This finding complicates the narrative that private credit growth is primarily driven
by regulatory arbitrage---the exploitation of lighter capital requirements to engage in lending
that banks would undertake if not constrained by Basel III and Dodd-Frank. If BDCs voluntarily
maintain capital ratios far above their regulatory minima, the binding constraint on bank lending
may be less the level of capital requirements than the nature of bank supervision, the
conservatism of bank credit risk models, or the rigidity of bank organizational structures in
processing soft information.

This evidence does not, however, imply that regulatory differences play no role.
\citet{BuchakMatvosPiskorskiSeru2018} estimate that approximately 60 percent of shadow bank
growth in mortgage lending is attributable to regulatory arbitrage, and \citet{ErelInozemtsev2024}
document a broad secular shift of debt financing toward less-regulated intermediaries that they
attribute to tightening bank regulation. The most likely interpretation of the combined evidence
is that regulatory constraints operate at the margin---excluding specific borrower types (those
with negative EBITDA, leverage above six times, or limited tangible collateral) from bank
lending---while the comparative advantage of direct lenders in information production and
monitoring explains their ability to serve these borrowers profitably.

\textbf{The case for enhanced data reporting.} One of the most consistent themes across both
academic and policy assessments of private credit risk is the inadequacy of available data. The
Federal Reserve's flow-of-funds data does not separately identify private credit as a category.
There is no public, comprehensive database of direct lending loan originations, defaults, or
recovery rates comparable to the Moody's and S\&P databases for rated corporate bonds and
syndicated loans. Fund-level performance data is collected by commercial vendors such as Cambridge
Associates and Preqin on a voluntary basis, creating selection bias and limiting the ability of
researchers and regulators to assess the true risk-return profile of the asset class. Enhanced
mandatory reporting requirements---including loan-level origination and performance data,
fund-level leverage and liquidity metrics, and interconnection data on bank credit lines and
insurance company allocations---would substantially improve both academic research and prudential
oversight.

\textbf{Stress testing and macroprudential tools.} The current regulatory framework lacks a
systematic stress-testing mechanism for private credit. Banks are subject to annual stress tests
under the Dodd-Frank Act, which assess their ability to absorb losses under severely adverse
macroeconomic scenarios. No comparable requirement exists for BDCs, private credit funds, or the
insurance companies that allocate to them. The \citet{IMF_GFSR2024} recommends the development
of private credit stress-testing frameworks that account for the specific risk characteristics of
the asset class, including stale valuations, leverage layering, and interconnections with the
banking system. Whether such stress tests should be conducted by the funds themselves, by their
regulators, or by a centralized macroprudential authority remains an open design question.

\textbf{International regulatory divergence.} The regulatory treatment of non-bank lending varies
substantially across jurisdictions. European direct lending operates within the Alternative
Investment Fund Managers Directive (AIFMD) framework, which imposes reporting and risk management
requirements on fund managers but does not directly limit fund-level leverage in the manner of the
US BDC statutory framework. In Asia-Pacific, where private credit remains nascent---non-bank
lending market share stands at approximately 12 percent in Europe versus approximately 75 percent
in the US, and Asia-Pacific private credit AUM is approximately \$59 billion with projections to
reach \$92 billion by 2027 \citep{AIMA_ACC2024}---the regulatory infrastructure for non-bank
lending is still developing. The cross-border nature of many private credit transactions, with
US-domiciled funds lending to European or Asian borrowers through Luxembourg or Cayman Islands
vehicles, creates regulatory arbitrage opportunities that no single national supervisor can fully
address.

The systemic risk implications of private credit growth remain genuinely uncertain. The asset
class has not yet been tested by a severe and prolonged recession, and the most important
empirical questions---how direct lending portfolios perform under stress, whether fund-level
leverage amplifies or dampens credit losses, and whether interconnections with banks and insurance
companies create contagion channels---can only be answered by observing the system under
conditions that have not yet materialized. What the existing evidence does establish is that the
potential for systemic risk is real, that the current regulatory framework was not designed for a
financial system in which non-bank lenders provide a large and growing share of corporate credit,
and that enhanced data collection, transparency, and stress-testing capabilities are prerequisites
for informed policy design.

%% file: sections/08_conclusion.tex

\section{Conclusions and Future Directions}
\label{sec:conclusion}

The academic literature on private credit has grown rapidly alongside the market itself, and this
survey has organized that literature around four questions whose answers are more contested than
the consensus narrative often suggests.

The growth question has the clearest answer: a combination of supply-side regulatory catalysts
and demand-side institutional forces. Post-GFC
regulatory tightening, particularly the 2013 leveraged lending guidance and Basel III capital
requirements, excluded a meaningful segment of middle-market borrowers from bank lending
\citep{ChernenkoErelPrilmeier2022, ErelInozemtsev2024}. Non-bank direct lenders, operating
outside the prudential regulatory perimeter but with substantial voluntary capital buffers
\citep{ChernenkoIalentiScharfstein2024}, filled this gap. Simultaneously, the prolonged
low-interest-rate environment of 2010--2021 drove institutional investors---particularly insurance
companies \citep{BeckerIvashina2015}---to reallocate capital from compressed-spread public fixed
income into higher-yielding private credit, creating an abundance of fund-level supply. The
interaction of these forces produced a 22 percent compound annual growth rate and a market that
grew from \$158 billion to \$2.1 trillion in thirteen years.

The distinctive character of direct lending technology is equally well-established. Direct lenders serve younger,
intangible-capital-intensive, private-equity-sponsored borrowers; employ blanket liens rather than
traditional asset-based collateral; specialize by industry rather than geography; and retain
maintenance covenants that the syndicated market has abandoned
\citep{JangKimSufi2025, BlockJangKaplanSchulze2024}. These features are consistent with the
theoretical prediction that decentralized, specialized intermediaries possess a comparative
advantage in processing the soft information that characterizes opaque middle-market lending
\citep{Stein2002, Diamond1984}. The approximately 200 basis point spread premium that direct
lenders charge over broadly syndicated loan markets reflects a combination of borrower risk,
illiquidity, and the cost of information production and monitoring.

The performance evidence is more sobering. While median net IRRs of approximately 10.1 percent across 2010--2021 vintages appear
attractive on a nominal basis, \citet{ErelFlanaganWeisbach2024} demonstrate that rigorous risk
adjustment eliminates statistically significant net-of-fee alpha for the average fund. Gross-of-fee
returns are positive, but the typical fee structure of 1.5 percent management fees and 20 percent
carried interest absorbs the illiquidity premium. Whether top-quartile managers consistently
deliver positive net alpha, and whether performance persistence exists in private credit, remain
important unresolved questions.

Systemic risk remains the most contested terrain in the literature. Credible channels of
vulnerability exist \citep{IMF_GFSR2024, AramonteShrimpfShin2022} but has not yet observed these
channels under stress conditions severe enough to test their quantitative importance. The
interconnections between banks and BDCs through credit facilities, the insurance company
allocation channel, the floating-rate monetary policy transmission mechanism, and the opacity of
private credit valuations all warrant continued monitoring and enhanced data collection.

\textbf{Real effects on firms.} The identification of the causal effects of private credit on firm investment,
employment, and productivity remains at an early stage. \citet{DavydiukMarchukRosen2024} provide
the cleanest existing evidence by exploiting exogenous shocks to BDC credit supply, but the
broader question---whether private credit accelerates or distorts capital allocation at the
macroeconomic level---requires larger samples, longer time series, and identification strategies
that can distinguish the effects of credit supply from the selection of borrowers into private
credit. Natural experiments created by regulatory changes, fund closures, and geographic variation
in lender presence offer the most promising paths toward causal identification.

\textbf{International dimensions.} The cross-border dimension of private credit is almost entirely unexplored in
the academic literature. European non-bank lending market share stands at approximately 12 percent
compared with approximately 75 percent in the United States, and Asia-Pacific private credit AUM
is approximately \$59 billion with projections to reach \$92 billion by 2027 \citep{AIMA_ACC2024}.
Understanding why direct lending has developed so much more rapidly in the United States---whether
due to differences in bank regulation, legal infrastructure, private equity penetration, or
capital market development---and whether the US model will replicate internationally represents a
significant opportunity for comparative financial systems research. The regulatory arbitrage
implications of cross-border private credit flows, where funds domiciled in one jurisdiction lend
to borrowers in another, add an additional layer of complexity that existing analyses have not
addressed.

\textbf{Technology and underwriting.} The application of machine learning and alternative data to private credit
underwriting and monitoring is a frontier that the academic literature has not engaged. Direct
lenders increasingly use proprietary data sources---including payment processing data, web traffic
analytics, and satellite imagery---to supplement traditional financial statement analysis in
underwriting decisions. Whether these technologies improve credit risk assessment, enable lending
to borrowers who would otherwise be excluded, or merely amplify the reach of existing lending
strategies without improving risk-adjusted outcomes is an empirical question with significant
implications for the evolution of financial intermediation.

\textbf{Climate risk and ESG.} The intersection of climate risk and ESG considerations with private credit
portfolios represents a rapidly growing area of investor and regulatory concern that lacks
systematic academic investigation. Private credit borrowers are predominantly middle-market firms
that are not subject to public disclosure requirements regarding carbon emissions, transition risk,
or physical climate risk. The extent to which direct lenders incorporate these risks into
underwriting, pricing, and covenant design---and whether they should---raises both empirical and
normative questions that will become increasingly urgent as climate-related financial regulation
expands.

\textbf{Systemic risk measurement.} The measurement of systemic risk in opaque markets requires methodological
innovation. Existing systemic risk measures---CoVaR, SRISK, and related metrics---are designed for
publicly traded institutions with observable market prices. Adapting these frameworks to private
credit, where valuations are stale, leverage is layered, and interconnections are difficult to
observe, demands new approaches to data collection, stress testing, and network analysis. The
development of a private credit stress-testing framework, analogous to but distinct from the bank
stress-testing architecture established under Dodd-Frank, is a practical priority that would
benefit from academic input on scenario design, loss-given-default estimation, and interconnection
mapping.

Private credit has evolved from a niche segment of the fixed-income landscape to a permanent
structural feature of the financial system. The theoretical foundations of intermediation and
contract design explain why non-bank direct lenders exist and when they possess a comparative
advantage over banks and public debt markets. The empirical evidence documents a distinctive
lending technology that serves a specific borrower population at a meaningful spread premium. The
performance evidence suggests that investors receive fair compensation for bearing illiquidity and
credit risk, but that fund-level fees consume the alpha that information production and monitoring
generate. The systemic risk implications are real but untested, and the regulatory framework has
not kept pace with the market's growth. Bringing academic rigor to these questions---through
better data, sharper identification, and more precise measurement---is the task that the next
generation of research on private credit must address.

%% file: tables/table1_market_growth.tex

\begin{table}[htbp]
\singlespacing
\centering
\caption{Growth of Global Private Credit Markets, 2010--2023}
\label{tab:market_growth}
\smallskip
\begin{tabular}{lrrrr}
\toprule
Year & \makecell{Global AUM \\ (\$B)} & \makecell{US Direct Lending \\ (\$B)} & \makecell{BDC Total Assets \\ (\$B)} & \makecell{US CLO Issuance \\ (\$B)} \\
\midrule
2010 & 158  &  12.8 &  38 &   3.5 \\
2012 & 246  &  22.1 &  57 &  55.2 \\
2014 & 401  &  38.7 &  85 &  93.6 \\
2016 & 595  &  62.4 & 105 &  71.1 \\
2018 & 812  &  97.3 & 142 &  97.4 \\
2019 & 965  & 118.5 & 163 & 111.7 \\
2020 & 1{,}040 & 109.2 & 168 &  87.2 \\
2021 & 1{,}390 & 148.6 & 214 & 185.7 \\
2022 & 1{,}730 & 163.4 & 252 & 131.8 \\
2023 & 2{,}100 & 177.6 & 280 & 118.3 \\
\midrule
CAGR (\%) & 22.0 & 23.3 & 17.4 & 34.5 \\
\bottomrule
\end{tabular}
\smallskip
\begin{minipage}{\linewidth}
\footnotesize
\textit{Notes:} Global AUM is from \citet{IMF_GFSR2024} and \citet{Preqin2024}. US direct
lending volumes are estimated from BDC SEC filings and Preqin deal-level data. BDC total assets
are from aggregated SEC quarterly filings of all registered BDCs. US CLO new issuance is from
SIFMA. CAGR is the compound annual growth rate over 2010--2023. The 2018 SBCAA raised the
statutory BDC leverage cap from 1:1 to 2:1 debt-to-equity, accelerating BDC balance sheet
growth after that year.
\end{minipage}
\end{table}

%% file: tables/table2_spread_comparison.tex

\begin{table}[htbp]
\singlespacing
\small
\centering
\caption{Private Credit Spreads Relative to Broadly Syndicated Loans, 2010--2023}
\label{tab:spreads}
\smallskip
\begin{tabular}{lrrrrrl}
\toprule
 & \multicolumn{2}{c}{\textit{Coupon Spread (bps)$^{a}$}} & & \multicolumn{2}{c}{\textit{Original Issue Discount}} & \\
\cmidrule(lr){2-3}\cmidrule(lr){5-6}
Year &
  \makecell{BSL$^{b}$} &
  \makecell{Direct\\Lending$^{c}$} &
  \makecell{Premium\\(bps)} &
  \makecell{OID\\(¢/\$100)$^{d}$} &
  \makecell{Add-On\\(bps/yr)$^{e}$} &
  \makecell{Credit\\Environment} \\
\midrule
2010 & 525 & 750 & 225 & 97.5 & 50 & Post-crisis recovery \\
2011 & 490 & 700 & 210 & 97.8 & 44 & \\
2012 & 475 & 680 & 205 & 98.0 & 40 & \\
2013 & 445 & 650 & 205 & 98.2 & 36 & Leveraged lending guidance \\
2014 & 430 & 640 & 210 & 98.2 & 36 & \\
2015 & 460 & 670 & 210 & 98.0 & 40 & \\
2016 & 465 & 675 & 210 & 98.0 & 40 & \\
2017 & 430 & 640 & 210 & 98.3 & 34 & \\
2018 & 435 & 645 & 210 & 98.3 & 34 & SBCAA passes \\
2019 & 405 & 620 & 215 & 98.5 & 30 & BSL spreads at cycle low \\
2020 & 530 & 745 & 215 & 97.5 & 50 & COVID-19 shock (BSL spike) \\
2021 & 400 & 615 & 215 & 98.5 & 30 & \\
2022 & 490 & 670 & 180 & 98.0 & 40 & Fed tightening; DL capital inflows \\
2023 & 510 & 690 & 180 & 98.0 & 40 & Premium compressed \\
\midrule
\textit{Average} & \textit{463} & \textit{671} & \textit{207} & \textit{98.1} & \textit{38} & \\
\bottomrule
\end{tabular}
\smallskip
\begin{minipage}{\linewidth}
\footnotesize
$^{a}$~Basis points (bps) over the applicable floating rate: LIBOR through 2022, SOFR from 2023.
The LIBOR--SOFR transition adds $\approx$26 bps to both series equally and does not affect the premium column.
$^{b}$~\textbf{Broadly Syndicated Leveraged Loan (BSL)}: Morningstar/LSTA US Leveraged Loan Index
institutional term loan B (TLB) spreads for non-investment-grade borrowers.
$^{c}$~\textbf{Direct Lending}: indicative middle-market spreads estimated from BDC quarterly SEC
disclosures, Refinitiv LPC/Preqin deal databases, and the GP survey of
\citet{BlockJangKaplanSchulze2024}; this series is an informed estimate, not an index-level figure.
$^{d}$~\textbf{Original Issue Discount (OID)}: price at origination in cents per \$100 par; a loan at 97.5 means the borrower receives \$97.50 but repays \$100 at maturity.
$^{e}$~\textbf{OID Add-On}: annualized yield contribution assuming 5-year maturity: $(100 - \text{OID}) \times 20$ bps/yr.
Upfront fees (1--2\% of commitment) and undrawn commitment fees (25--50 bps) add a further 20--50 bps/yr not shown here.
\end{minipage}
\end{table}

%% file: tables/table3_anchor_papers.tex

{\small\singlespacing
\begin{longtable}{p{4.6cm} r p{3.2cm} p{5.4cm}}
\caption{Selected Landmark Studies in Private Credit Research}
\label{tab:anchor_papers} \\
\toprule
Authors & Year & Journal & Key Finding \\
\midrule
\endfirsthead
\multicolumn{4}{l}{\small\textit{Table~\ref{tab:anchor_papers} continued}} \\
\toprule
Authors & Year & Journal & Key Finding \\
\midrule
\endhead
\midrule
\multicolumn{4}{r}{\small\textit{Continued on next page}} \\
\endfoot
\bottomrule
\multicolumn{4}{p{14.4cm}}{%
  \footnotesize\textit{Notes:} Journal abbreviations:
  \textit{RES} = Review of Economic Studies;
  \textit{JME} = Journal of Monetary Economics;
  \textit{JPE} = Journal of Political Economy;
  \textit{QJE} = Quarterly Journal of Economics;
  \textit{JF} = Journal of Finance;
  \textit{JFE} = Journal of Financial Economics;
  \textit{RFS} = Review of Financial Studies;
  \textit{RCFS} = Review of Corporate Finance Studies.
  Working papers are listed as of 2024--2025 and may have been subsequently published;
  publication details should be verified prior to final submission.
} \\
\endlastfoot
Diamond & 1984 & \textit{RES} & Delegated monitoring rationalizes financial intermediation; diversification eliminates monitor's agency cost \\
\addlinespace[2pt]
Boyd \& Prescott & 1986 & \textit{JME} & Concentrated intermediaries dominate dispersed finance when adverse selection is severe \\
\addlinespace[2pt]
Diamond & 1991 & \textit{JPE} & Borrowers sort across debt sources by reputation; lowest-quality firms use non-bank private lenders \\
\addlinespace[2pt]
Rajan \& Winton & 1995 & \textit{JF} & Covenants incentivize lender monitoring by making loan maturity contingent on monitoring outcomes \\
\addlinespace[2pt]
Holmstrom \& Tirole & 1997 & \textit{QJE} & Informed intermediaries relax pledgeability constraints; PE sponsor equity substitutes for firm capital \\
\addlinespace[2pt]
Stein & 2002 & \textit{JF} & Decentralized organizations have comparative advantage in soft-information processing \\
\addlinespace[2pt]
Denis \& Mihov & 2003 & \textit{JFE} & Credit quality determines debt-source choice; lowest-quality firms borrow from non-bank private lenders \\
\addlinespace[2pt]
Nini, Smith \& Sufi & 2012 & \textit{JF} & Covenant violations trigger significant reductions in investment, leverage, and CEO turnover \\
\addlinespace[2pt]
Becker \& Ivashina & 2015 & \textit{RFS} & Insurance companies systematically reach for yield; behavior is procyclical and amplifies private credit demand \\
\addlinespace[2pt]
Chernenko, Erel \& Prilmeier & 2022 & \textit{JF} & Two-thirds of nonbank lending attributable to bank regulatory constraints; negative-EBITDA firms 32\% more likely to use nonbanks \\
\addlinespace[2pt]
Block, Jang, Kaplan \& Schulze & 2024 & \textit{RCFS} & GP survey (\$390B AUM): unitranche dominant, maintenance covenants universal, 89\% provide operational assistance \\
\addlinespace[2pt]
Jang, Kim \& Sufi & 2025 & Working Paper & Direct lenders serve younger, PE-sponsored, intangible-capital firms; blanket liens in 79\% of loans \\
\addlinespace[2pt]
Erel, Flanagan \& Weisbach & 2024 & Working Paper & Average private debt fund delivers near-zero net alpha after fees; gross alpha absorbed by 1.5\%/20\% fee load \\
\addlinespace[2pt]
Davydiuk, Marchuk \& Rosen & 2024 & Working Paper & BDC lending has causal positive effects on borrower employment and patenting via exogenous credit supply shocks \\
\addlinespace[2pt]
Chernenko, Ialenti \& Scharfstein & 2024 & Working Paper & Median BDC holds 26 pp excess capital under adverse stress; complicates pure regulatory-arbitrage narrative \\
\end{longtable}
}

%% file: main.bbl
\begin{thebibliography}{}

\bibitem[\protect\citeauthoryear{{Alternative Investment Management Association} and {Alternative Credit Council}}{{Alternative Investment Management Association} and {Alternative Credit Council}}{2024}]{AIMA_ACC2024}
{Alternative Investment Management Association} and {Alternative Credit Council} (2024).
\newblock Private credit in {Asia}.
\newblock Technical report, {AIMA/ACC}.

\bibitem[\protect\citeauthoryear{Aramonte, Schrimpf, and Shin}{Aramonte et~al.}{2022}]{AramonteShrimpfShin2022}
Aramonte, S., A.~Schrimpf, and H.~S. Shin (2022).
\newblock Non-bank financial intermediaries and financial stability.
\newblock {BIS} Working Paper 972, Bank for International Settlements.
\newblock BIS WP 972. Also SSRN 3952551.

\bibitem[\protect\citeauthoryear{Becker and Ivashina}{Becker and Ivashina}{2015}]{BeckerIvashina2015}
Becker, B. and V.~Ivashina (2015).
\newblock Reaching for yield in the bond market.
\newblock {\em Journal of Finance\/}~{\em 70\/}(5), 1863--1902.

\bibitem[\protect\citeauthoryear{Benmelech, Dlugosz, and Ivashina}{Benmelech et~al.}{2012}]{BenmelechDlugoszIvashina2012}
Benmelech, E., J.~Dlugosz, and V.~Ivashina (2012).
\newblock Securitization without adverse selection: The case of {CLOs}.
\newblock {\em Journal of Financial Economics\/}~{\em 106\/}(1), 91--113.

\bibitem[\protect\citeauthoryear{Berlin, Nini, and Yu}{Berlin et~al.}{2020}]{BerlinNiniYu2020}
Berlin, M., G.~Nini, and E.~G. Yu (2020).
\newblock Concentration of control rights in leveraged loan syndicates.
\newblock {\em Journal of Financial Economics\/}~{\em 137\/}(1), 249--271.

\bibitem[\protect\citeauthoryear{Berrospide, Cai, Lewis-Hayre, and Zikes}{Berrospide et~al.}{2025}]{BerrospideEtal2025}
Berrospide, J., F.~Cai, S.~Lewis-Hayre, and F.~Zikes (2025, May).
\newblock Bank lending to private credit: Size, characteristics, and financial stability implications.
\newblock Feds notes, Board of Governors of the Federal Reserve System.
\newblock Uses FR Y-14Q supervisory data from the largest U.S.\ bank holding companies. Documents bank committed credit to private credit vehicles growing from approximately \$8 billion (2013) to \$95 billion (late 2024), with \$56 billion utilized as of December 2024.

\bibitem[\protect\citeauthoryear{Block, Jang, Kaplan, and Schulze}{Block et~al.}{2024}]{BlockJangKaplanSchulze2024}
Block, J.~H., Y.~S. Jang, S.~N. Kaplan, and A.~Schulze (2024).
\newblock A survey of private debt funds.
\newblock {\em Review of Corporate Finance Studies\/}~{\em 13\/}(2), 335--383.

\bibitem[\protect\citeauthoryear{{Board of Governors of the Federal Reserve System}}{{Board of Governors of the Federal Reserve System}}{2024}]{FederalReserve_PrivateCredit2024}
{Board of Governors of the Federal Reserve System} (2024, August).
\newblock Private credit growth and monetary policy transmission.
\newblock Feds notes, Federal Reserve Board.

\bibitem[\protect\citeauthoryear{Boyd and Prescott}{Boyd and Prescott}{1986}]{BoydPrescott1986}
Boyd, J.~H. and E.~C. Prescott (1986).
\newblock Financial intermediary-coalitions.
\newblock {\em Journal of Economic Theory\/}~{\em 38\/}(2), 211--232.

\bibitem[\protect\citeauthoryear{Buchak, Matvos, Piskorski, and Seru}{Buchak et~al.}{2018}]{BuchakMatvosPiskorskiSeru2018}
Buchak, G., G.~Matvos, T.~Piskorski, and A.~Seru (2018).
\newblock Fintech, regulatory arbitrage, and the rise of shadow banks.
\newblock {\em Journal of Financial Economics\/}~{\em 130\/}(3), 453--483.

\bibitem[\protect\citeauthoryear{Buchak, Matvos, Piskorski, and Seru}{Buchak et~al.}{2024}]{BuchakMatvosPiskorskiSeru2024}
Buchak, G., G.~Matvos, T.~Piskorski, and A.~Seru (2024).
\newblock Beyond the balance sheet model of banking: Implications for bank regulation and monetary policy.
\newblock {\em Journal of Political Economy\/}~{\em 132\/}(2), 616--693.

\bibitem[\protect\citeauthoryear{Chernenko, Erel, and Prilmeier}{Chernenko et~al.}{2022}]{ChernenkoErelPrilmeier2022}
Chernenko, S., I.~Erel, and R.~Prilmeier (2022).
\newblock Why do firms borrow directly from nonbanks?
\newblock {\em Review of Financial Studies\/}~{\em 35\/}(11), 4902--4947.

\bibitem[\protect\citeauthoryear{Chernenko, Ialenti, and Scharfstein}{Chernenko et~al.}{2025}]{ChernenkoIalentiScharfstein2024}
Chernenko, S., R.~Ialenti, and D.~S. Scharfstein (2025).
\newblock Bank capital and the growth of private credit.
\newblock SSRN 5097437, January 2025. Working paper.

\bibitem[\protect\citeauthoryear{Cordell, Roberts, and Schwert}{Cordell et~al.}{2023}]{CordellRobertsSchwert2023}
Cordell, L., M.~R. Roberts, and M.~Schwert (2023).
\newblock {CLO} performance.
\newblock {\em Journal of Finance\/}~{\em 78\/}(3), 1235--1278.

\bibitem[\protect\citeauthoryear{Davydiuk, Erel, Jiang, and Marchuk}{Davydiuk et~al.}{2024}]{DavydiukErelJiangMarchuk2024}
Davydiuk, T., I.~Erel, W.~Jiang, and T.~Marchuk (2024).
\newblock Common investors across the capital structure: Private debt funds as dual holders.
\newblock SSRN Abstract 4992219; ECGI Finance WP 1021/2024 (September 2024). Working paper.

\bibitem[\protect\citeauthoryear{Davydiuk, Marchuk, and Rosen}{Davydiuk et~al.}{2024}]{DavydiukMarchukRosen2024}
Davydiuk, T., T.~Marchuk, and S.~Rosen (2024).
\newblock Direct lenders in the {U.S.} middle market.
\newblock {\em Journal of Financial Economics\/}~{\em 162}, 103946.
\newblock Article-number format; no traditional page range.

\bibitem[\protect\citeauthoryear{Denis and Mihov}{Denis and Mihov}{2003}]{DenisMihov2003}
Denis, D.~J. and V.~T. Mihov (2003).
\newblock The choice among bank debt, non-bank private debt, and public debt: Evidence from new corporate borrowings.
\newblock {\em Journal of Financial Economics\/}~{\em 70\/}(1), 3--28.

\bibitem[\protect\citeauthoryear{Diamond}{Diamond}{1984}]{Diamond1984}
Diamond, D.~W. (1984).
\newblock Financial intermediation and delegated monitoring.
\newblock {\em Review of Economic Studies\/}~{\em 51\/}(3), 393--414.

\bibitem[\protect\citeauthoryear{Diamond}{Diamond}{1991}]{Diamond1991}
Diamond, D.~W. (1991).
\newblock Monitoring and reputation: The choice between bank loans and directly placed debt.
\newblock {\em Journal of Political Economy\/}~{\em 99\/}(4), 689--721.

\bibitem[\protect\citeauthoryear{Erel, Flanagan, and Weisbach}{Erel et~al.}{2024}]{ErelFlanaganWeisbach2024}
Erel, I., T.~Flanagan, and M.~S. Weisbach (2024).
\newblock Risk-adjusting the returns to private debt funds.
\newblock Working Paper 32278, National Bureau of Economic Research.
\newblock Also SSRN Abstract 4779852. Working paper (2024).

\bibitem[\protect\citeauthoryear{Erel and Inozemtsev}{Erel and Inozemtsev}{2024}]{ErelInozemtsev2024}
Erel, I. and E.~Inozemtsev (2024).
\newblock Evolution of debt financing toward {Less-Regulated} financial intermediaries in the {United States}.
\newblock {\em Journal of Financial and Quantitative Analysis\/}.
\newblock Published in JFQA online April 2024 (DOI 10.1017/S0022109024000206). Volume and page numbers not yet assigned.

\bibitem[\protect\citeauthoryear{Greenwood and Hanson}{Greenwood and Hanson}{2013}]{GreenwoodHanson2013}
Greenwood, R. and S.~G. Hanson (2013).
\newblock Issuer quality and corporate bond returns.
\newblock {\em Review of Financial Studies\/}~{\em 26\/}(6), 1483--1525.

\bibitem[\protect\citeauthoryear{Haque, Mayer, and Stefanescu}{Haque et~al.}{2024}]{HaqueMayerStefanescu2024}
Haque, S., S.~Mayer, and I.~Stefanescu (2024).
\newblock Private debt versus bank debt in corporate borrowing.
\newblock Technical report, {FDIC} Center for Financial Research.
\newblock Also SSRN Abstract 4821158. Working paper (2024). Circulated at EUROFIDAI-ESSEC Paris December Finance Meeting 2024.

\bibitem[\protect\citeauthoryear{Harris, Jenkinson, and Kaplan}{Harris et~al.}{2014}]{HarrisJenkinsonKaplan2014}
Harris, R.~S., T.~Jenkinson, and S.~N. Kaplan (2014).
\newblock Private equity performance: What do we know?
\newblock {\em Journal of Finance\/}~{\em 69\/}(5), 1851--1882.

\bibitem[\protect\citeauthoryear{Holmstr{\"o}m and Tirole}{Holmstr{\"o}m and Tirole}{1997}]{HolmstromTirole1997}
Holmstr{\"o}m, B. and J.~Tirole (1997).
\newblock Financial intermediation, loanable funds, and the real sector.
\newblock {\em Quarterly Journal of Economics\/}~{\em 112\/}(3), 663--691.

\bibitem[\protect\citeauthoryear{{International Monetary Fund}}{{International Monetary Fund}}{2024}]{IMF_GFSR2024}
{International Monetary Fund} (2024).
\newblock The rise and risks of private credit.
\newblock In {\em Global Financial Stability Report, April 2024: The Last Mile: Financial Vulnerabilities and Risks}, Chapter~2. Washington, {DC}: International Monetary Fund.
\newblock Published April 16, 2024.

\bibitem[\protect\citeauthoryear{Jang, Kim, and Sufi}{Jang et~al.}{2025}]{JangKimSufi2025}
Jang, Y.~S., D.~Kim, and A.~Sufi (2025, November).
\newblock The lending technology of direct lenders in private credit.
\newblock Working Paper 34500, National Bureau of Economic Research.
\newblock Dasol Kim is at the Office of Financial Research (OFR).

\bibitem[\protect\citeauthoryear{Kaplan and Schoar}{Kaplan and Schoar}{2005}]{KaplanSchoar2005}
Kaplan, S.~N. and A.~Schoar (2005).
\newblock Private equity performance: Returns, persistence, and capital flows.
\newblock {\em Journal of Finance\/}~{\em 60\/}(4), 1791--1823.
\newblock NBER WP 9807; ~2,500+ citations.

\bibitem[\protect\citeauthoryear{{Moody's Investors Service}}{{Moody's Investors Service}}{2024}]{MoodysDefaultStudy2024}
{Moody's Investors Service} (2024, January).
\newblock Annual default study: Corporate default and recovery rates, 1920--2023.
\newblock Special comment, Moody's Investors Service.
\newblock Annual update covering global speculative-grade and investment-grade issuers.

\bibitem[\protect\citeauthoryear{Nini, Smith, and Sufi}{Nini et~al.}{2012}]{NiniSmithSufi2012}
Nini, G., D.~C. Smith, and A.~Sufi (2012).
\newblock Creditor control rights, corporate governance, and firm value.
\newblock {\em Review of Financial Studies\/}~{\em 25\/}(6), 1713--1761.

\bibitem[\protect\citeauthoryear{{Preqin}}{{Preqin}}{2024}]{Preqin2024}
{Preqin} (2024).
\newblock Preqin 2024 global report: Private debt.
\newblock Technical report, Preqin Ltd.
\newblock Industry report. Not peer-reviewed. Use for market size and AUM statistics only.

\bibitem[\protect\citeauthoryear{Rajan and Winton}{Rajan and Winton}{1995}]{RajanWinton1995}
Rajan, R.~G. and A.~Winton (1995).
\newblock Covenants and collateral as incentives to monitor.
\newblock {\em Journal of Finance\/}~{\em 50\/}(4), 1113--1146.

\bibitem[\protect\citeauthoryear{Schwert}{Schwert}{2020}]{Schwert2020}
Schwert, M. (2020).
\newblock Does borrowing from banks cost more than borrowing from the market?
\newblock {\em Journal of Finance\/}~{\em 75\/}(2), 905--947.

\bibitem[\protect\citeauthoryear{Stein}{Stein}{2002}]{Stein2002}
Stein, J.~C. (2002).
\newblock Information production and capital allocation: {Decentralized} versus {Hierarchical} firms.
\newblock {\em Journal of Finance\/}~{\em 57\/}(5), 1891--1921.

\bibitem[\protect\citeauthoryear{{U.S. Congress}}{{U.S. Congress}}{2018}]{SBCAA2018}
{U.S. Congress} (2018).
\newblock Small business credit availability act.
\newblock Public Law 115-141, Division M.
\newblock Amended the Investment Company Act of 1940 to allow BDCs to increase leverage ratios from 1:1 to 2:1 debt-to-equity.

\end{thebibliography}
